# Rotational Coherence of Encapsulated Ortho and Para Water in Fullerene-$C_{60}$

Sergey S. Zhukov[1†], Vasileios Balos[2†], Gabriela Hoffman[3], Shamim Alom[3], Mikhail Belyanchikov[1], Mehmet Nebioglu[4], Seulki Roh[4], Artem Pronin[4], George R. Bacanu[3], Pavel Abramov[1], Martin Wolf[2], Martin Dressel[1,4], Malcolm H. Levitt[3], Richard J. Whitby[3], Boris Gorshunov[1*], Mohsen Sajadi[2*]

[1]Moscow Institute of Physics and Technology, Moscow, Russia.

[2]Fritz-Haber-Institut der MPG, Berlin, Germany.

[3]School of Chemistry, University of Southampton, Southampton, United Kingdom.

[4]1.Physikalisches Institut, Universität Stuttgart, Stuttgart, Germany.

*bpgorshunov@gmail.com, sajadi@fhi-berlin.mpg.de.

†These authors contributed equally in this work.

*Abstract*. Encapsulation of a single water molecule in fullerene-$C_{60}$ via chemical surgery provides a unique opportunity to study the distinct rotational dynamics of the water spin isomers at cryogenic temperatures. Here, we employ single-cycle terahertz (THz) pulses to coherently excite the low-frequency rotational motion of ortho- and para-water, encapsulated in fullerene-$C_{60}$. The THz pulse slightly orients the water electric dipole moments along the field polarization leading to the subsequent emission of electromagnetic waves, which we resolve via the field-free electro-optic sampling technique. At temperatures above ~100 K, the rotation of water in its cage is overdamped and no emission is resolved. At lower temperatures, the water rotation gains a long coherence decay time, allowing observation of the coherent emission for 10-15 ps after the initial excitation. We observe the real-time change of the emission pattern after cooling to 4 K, corresponding to the conversion of a mixture of ortho-water to para-water over the course of 10 hours.

Water is arguably the most important molecule for the life on earth. Its numerous physical, chemical and thermodynamic properties stimulated many studies in the past decades [1,2]. To unravel the complex behavior of water, single molecule rotational spectroscopies have drawn considerable attention [3–8]. An intriguing aspect of water dynamics is the distinct rotational quantum behavior of its ortho ($I=1$) and para ($I=0$) spin isomers. The nuclear spin isomerism of ortho- and para-water and its associated dynamics are also relevant to diverse scientific disciplines including magnetic resonance imaging [9], nano-electronics [10], quantum computing [11], astrophysics [12,13] and spin chemistry [14,15]. Therefore, significant efforts have been undertaken to separate and study ortho- and para-water.

Despite some strongly disputed claims [16], the physical separation of the spin isomers of bulk water is not thought to be feasible [17,18]. Kravchuk et al. formed a magnetically focused molecular beam of ortho-water, based on the deflection of the nuclear magnetic moments [19] and Horke et al. used strong static electric fields to gain spatial separation of the two isomers of water in a molecular beam [17,20]. Despite these salient achievements, fast proton transfer in bulk water limits the implementation of these approaches for gaining insight into the physical properties of ortho- and para-water.

Kurotobi et al. made a breakthrough in water research by encapsulating isolated water molecules in fullerene-$C_{60}$ cages, a highly symmetrical, homogenous and isolated environment [21]. The $H_2O@C_{60}$ system provides a unique opportunity to study the distinct rotational quantum dynamics of spin isomers of water at cryogenic temperatures without freezing its rotational motion. So far, the conversion of ortho-water to para-water was studied below 5 K, using various methods including NMR [22,23], mid-IR [24], FIR [21], inelastic neutron scattering (INS) [21,25] and dielectric measurements [26]. The INS also revealed the splitting of the triply-degenerate ortho-$H_2O$ rotational ground state into a singlet and a doublet state [21,22].

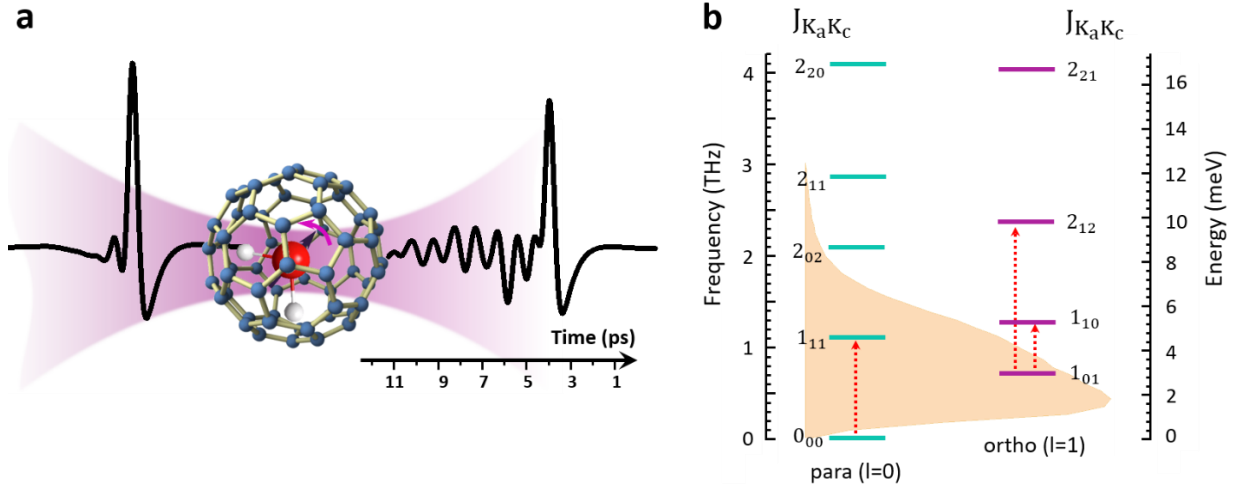

**Fig. 1**. **THz time-domain spectroscopy of $H_2O@C_{60}$. a**, A single cycle-THz pulse traverses the $H_2O@C_{60}$ sample and launches a rotational wave packet in the rotational quantum states of encapsulated water molecules. The coherent radiation of the molecules is observed as the tail of the transmitted THz pulse. **b**, The lowest rotational energy levels of water molecules. The para and ortho energy levels are labelled separately by the quantum numbers $J_{K_aK_c}$ (see text for details). The amber-colored area shows the frequency spectrum of the THz pump pulse. The arrows indicate the three lowest rotational transitions of water.

The experimental data [20–23] indicate that the lowest rotational states of water are almost unperturbed by the encapsulating cage at low temperature. The high symmetry of the cage allows water to rotate freely, with no perceptible hindering potential. This conclusion has been corroborated by nuclear magnetic resonance [27], molecular dynamics simulations [28], and multidimensional quantum calculations of the energy level structure of the encapsulated molecule [29,30]. Dielectric measurements of $H_2O@C_{60}$ indicate that the electric dipole moment of water is strongly screened by the cage, leading to a ~4-fold reduction in the effective dipole moment [23], in good agreement with quantum chemistry calculations [30]. The most prominent influence of the molecular environment on the lowest rotational states of the encapsulated water is the breaking of the 3-fold degeneracy of the ortho-water ground state [21,22], which has been attributed to an electric quadrupolar field generated by neighboring fullerene molecules in the lattice [31–33].

In this work, we measure the real-time rotational motion of water encapsulated inside $C_{60}$, by using time-domain terahertz (THz) spectroscopy. Because of the strong coupling between the THz electric-field and the permanent dipole moment of water, this method is very sensitive to the details of the rotational motion of encapsulated water inside its cage. As illustrated in **Fig. 1a**, upon THz excitation of the low-frequency rotational motions of water, the THz electric field $E(t)$ exerts a torque on the water's dipole moment $\mu$ via the interaction potential $H_1 = -\mathbf{\mu} \cdot \mathbf{E}(t) = -\mu\,E(t)\cos\theta$, where $\theta$ is the angle between the THz field polarization axis and $\mathbf{\mu}$. This interaction induces a coherent wave-packet in the rotational quantum states of water, such that one-quantum coherences (1QCs) are induced between the rotational eigenstates. The induced 1QCs cause the emission of electromagnetic waves whose amplitude decays exponentially, known as a free-induction-decay (FID) [34,35].

For a linear rotor, the rotational Hamiltonian is given by $\widehat{H}_0 = \frac{\widehat{J}^2}{2I}$, where $\hat{J}$ is the angular momentum operator and $I$ is the moment of inertia. Here, the molecular rotational transition frequencies are integer multiples of the lowest transition frequency, $\omega_{0\leftarrow 1} = 2Bc$ ($c$, the speed of light; $B$, the rotational constant). As the molecules are initially distributed in different $J$ states, the phase relation induced by the THz pulse is quickly lost and their FID emissions diphase and interfere destructively. After a period of $T_{\text{rev}} = (2Bc)^{-1}$, the revival period, the dipoles reorient and their FID emissions interfere constructively [31].

An asymmetric top molecule such as water has three different moments of inertia $I_a$, $I_b$ and $I_c$ along the three principal axes of the inertial tensor and the Hamiltonian is given by $\hat{H}_0 = \frac{\hat{J}_a^2}{2I_a} + \frac{\hat{J}_b^2}{2I_b} + \frac{\hat{J}_c^2}{2I_c}$. For this symmetry group, the molecular rotational motion gains a complex pattern, resembling that of a mixture of three distinct linear rotors. To label the rotational transitions of the asymmetric top molecules, we follow the procedure in which, each rotational energy level is denoted by the $J$ state and its projections, $K_a$ and $K_c$ along the principal axes $a$ and $c$ [36].

According to the Pauli principle, for water with identical fermionic protons, the total internal wave function is antisymmetric with respect to particle exchange [37]. For such system in its ground electronic and vibronic state, the wave function $|\psi\rangle = |\psi_{\text{rot}}\rangle|\psi_{\text{spin}}\rangle$ changes sign by the permutation of its hydrogen atoms. As a result, for para-water with antisymmetric nuclear-spin wave function $|\psi_{\text{spin}}\rangle$, the only allowed spatial wave functions are symmetric with respect to particle exchange. These are characterized by even values of $K_a+K_c$. In contrast, for ortho-water with symmetric $|\psi_{\text{spin}}\rangle$, the allowed spatial wave functions are antisymmetric with respect to particle exchange, corresponding to odd values of $K_a+K_c$. As depicted in **Fig. 1b**, the absolute rotational ground state of water belongs to para-water $J_{K_aK_c} = 0_{00}$ with $1_{11}$ and $1_{02}$ as its first and second excited rotational states. The ground rotational state of the ortho-water is $J_{K_aK_c} = 1_{01}$ and $1_{10}$ and $1_{12}$ are the first two excited rotational states.

In our THz experiment, we employ single-cycle terahertz (THz) pulses with duration of about 0.5 picosecond (ps) to excite the low-frequency rotational transitions of water. The amber colored area in **Fig. 1b**, shows the amplitude spectrum of the pulse. To resolve FID emissions, we employ the field-free electro-optic sampling technique [38], with temporal resolution <0.02 ps. The $H_2O@C_{60}$ sample is a bi-layer pellet, whose first layer (~0.2 mm thick) is a homogeneous mixture comprising 5% of $H_2O@C_{60}$ and 95% empty $C_{60}$ and the second layer is empty $C_{60}$ (~0.5 mm thick). The diluted 5% $H_2O@C_{60}$ is used to avoid saturation of the absorption transitions and the large thickness of the bi-layer pellet helps to delay the first reflected echo signal from the pellet-air interface to ~8 ps; opening a wide temporal window for capturing a large fraction of the water's FID emissions, without their interference with the reflected echo signal.

*Impact of temperature.* First, we study the impact of temperature, in the range of 6 K to 300 K, on the THz response of the encapsulated water. The THz response of $H_2O@C_{60}$, subtracted from that of the empty $C_{60}$ (see raw data before subtraction in Fig. S1) and their corresponding Fourier spectra are presented in **Fig. 2**. The onset of FID emissions can be resolved at temperatures below ~100 K, and by further cooling the sample, the emissions gain larger amplitude. Since $C_{60}$ is transparent at THz frequencies, the absence of FID emissions at temperatures above ~100 K is a clear indication of the strongly overdamped coherent rotational motion of encapsulated water inside its $C_{60}$ cage. In contrast, the coherent rotations of $H_2O$ in the vapor phase last for about ~100 ps at ambient temperature [39,40]. The damping effect of the environment on the water rotational dynamics can also be realized from the broadening of the rotational transitions. As shown in **Fig. 2c**, the full width at half maximum (FWHM) of the two strong bands at 1 THz and 1.53 THz increase strongly as the temperature is increased from 6 K to about 100 K. The rapid increase in linewidth for the ortho-$H_2O$ transition above ~80 K might be associated with a phase transition in $C_{60}$, with increased rotational freedom of the $C_{60}$ cages themselves above ~90 K [41].

By lowering the temperature, the encaged water molecules experience less fluctuations, hence the THz-induced rotational coherence of water lasts for a longer time. As a result, the FID emissions of water molecules can be resolved below T≈100 K. As shown in **Fig. 2b** and **Fig. 3a**, the emission encompasses different spectral components, with first three low-frequency contributions centered at ~0.52 THz, ~1 THz and ~1.53 THz, corresponding to the transitions with quantum energies 2.15 meV, 4.14 meV and 6.33 meV, respectively. The observed rotational transitions are in good agreement with the three lowest rotational transitions of free water [42]: the para transition $0_{00} \rightarrow 1_{11}$ with energy 4.56 meV, and two ortho transitions $1_{01} \rightarrow 1_{10}$ and $1_{01} \rightarrow 2_{12}$ with energies 2.26 meV and 6.85 meV (the three transitions are marked by red dashed arrows in **Fig. 1b**). The difference between the rotational frequencies of the free water and those of the encapsulated water, from the lowest to highest frequencies, are -1.2 cm⁻¹, +4 $cm^{-1}$ and +4.6 $cm^{-1}$, implying only a minor impact of the $C_{60}$ cage on the energy levels of the encapsulated water molecules, consistent with previous IFR and INS findings [21].

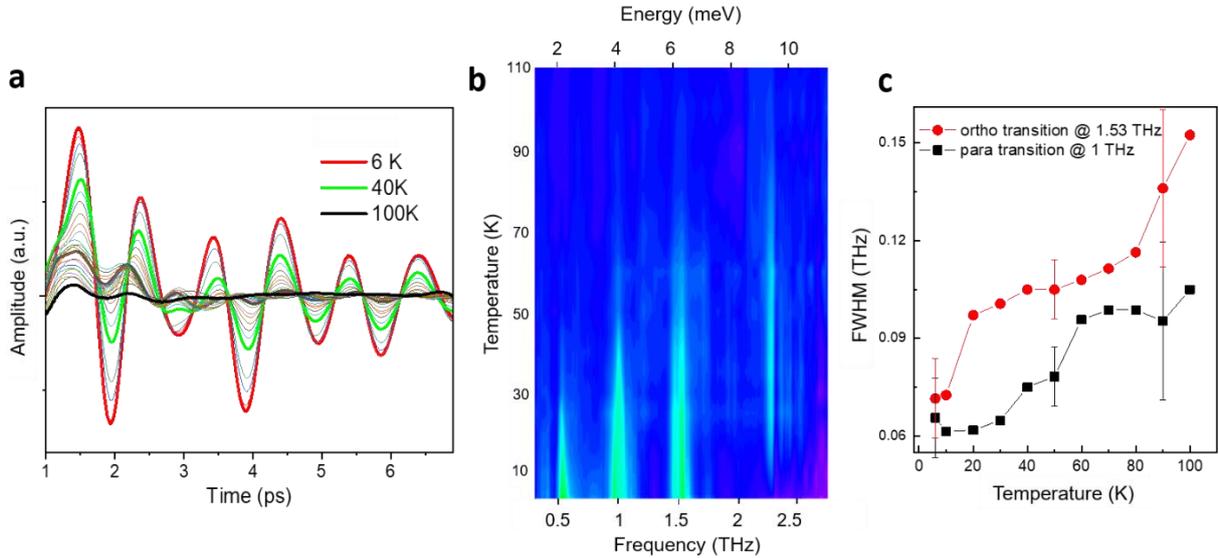

**Fig. 2**. **Temperature-dependent rotational motion of $H_2O@C_{60}$**. **a**, The rotational coherences of $H_2O$ are clearly resolved at cryogenic temperatures but display increasing damping as the temperature is increased. **b**, The contour plot of the Fourier spectra of the time-domain signals shows the rotational components (the bright lines) of the encapsulated water molecules as function of sample temperature. **c**, The rotational transitions at 1 THz and at 1.53 THz clearly show an increase in linewidth with increasing temperature.

In addition to the major three low-frequency rotational transitions, there are bright lines in the contour plot of **Fig. 2b** at frequencies higher than 1.5 THz. The selected spectra of the contour plot are displayed in **Fig. 3a** and show a relatively strong peak at 2.26 THz and two weak broad features, marked by asterisks, at about 1.9 THz and 2.49 THz. Moreover, as depicted in **Fig. 3b**, while the three low-frequency peaks (presented by the one at ~ 1THz) monotonically gain larger amplitude by lowering temperature, the peak at 2.26 THz gains its largest amplitude at about 50 K and completely vanishes below ~10 K. The fact that this transition only appears at higher temperatures strongly suggests that it originates from an excited rotational state. The selection rules for electromagnetic transitions are as follows: $\Delta J = 0, \pm 1, \Delta K_a = \pm 1, \Delta K_c = \pm 1$. Accordingly, the allowed transitions, originating from the first rotational excited states, are (the frequencies in isolated water molecules, from Ref [39] are given in brackets): *para* $1_{11} \rightarrow 2_{02}$ (0.99 THz), *para* $1_{11} \rightarrow 2_{11}$ (1.74 THz), *para* $1_{11} \rightarrow 2_{20}$ (2.97 THz) and *ortho* $1_{10} \rightarrow 2_{21}$ (2.77 THz).

None of these frequencies matches closely the observed 2.26 THz transition. The closest candidate is the ortho $1_{10} \rightarrow 2_{21}$ transition, but even this would require a red shift of ~15 $cm^{-1}$ with respect to free water. A red shift of this magnitude is far larger than those predicted by quantum calculations [25] and in contrast to the negligible difference between the three lowest energy levels of the free water and the encapsulated water. In the same way, the broad features at ~1.9 THz and ~2.49 THz would require, respectively a blue shift by ~ 30 $cm^{-1}$ and a red shift by ~ 15 $cm^{-1}$ to match the closest free water transition of $1_{11} \rightarrow 2_{02}$ and $1_{11} \rightarrow 2_{20}$.

At this point, the origin of the 2.26 THz peak is unknown. One possibility is that the potential energy function used for the existing quantum calculations [25,29,30] is insufficiently accurate to predict the behavior of the excited rotational states of the encapsulated water. Another possibility is that the 2.26 THz peak involves an interaction between the water rotation and the lattice phonons. The low-frequency phonons of the $C_{60}$ lattice, which are normally invisible to the THz radiation, might become visible due to their coupling to the water dipole. Neutron scattering studies have indicated the presence of low-frequency phonons in the $C_{60}$ lattice at about 2 THz [43].

*Ortho-para conversion.* Second, we measure the FID emissions of the encapsulated water at 4 K, over the course of ~30 hours. For these experiments, we gradually lowered the sample temperature to 20 K, and then rapidly lowered it to 4 K. The FID signals at waiting times equal to 1, 5 and 10 hours are shown in the left panel of **Fig. 4a**. An interesting observation is the gradual change of the FID emission pattern, from a more complex to a simpler pattern.

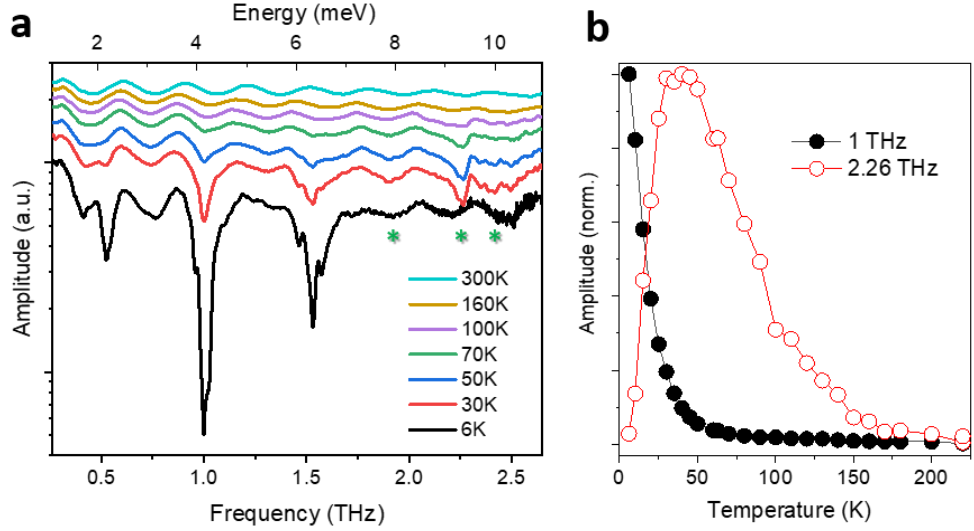

**Fig. 3**. **Temperature-dependent rotational spectra**. **a**, Selected Fourier spectra of the time-domain FID emissions of $H_2O@C_{60}$. The prominent peaks at 0.52 THz, 1 THz and 1.53 THz monotonically gain larger amplitude by lowering temperature and match well with rotational transitions of free water. The peak at 2.26 THz and two broad features at about 1.9 THz and 2.49 THz, marked by green asterisks, do not match the free water transitions (see text for details). The spectra are vertically displaced for clarity. Note the logarithmic scale of the vertical axis. Periodic oscillations are due to the Fabry–Pérot effect. **b**, The peak amplitude of the transitions at 1 THz and 2.26 THz, normalized against their maximum value.

We determine the FID emission components by fitting the measured signals with $\sum_i A_i\, e^{-\gamma t} \sin(2\pi v_i t)$, where $v_i$ and $\gamma$ are respectively, the frequency and the damping constant of the fitted components [44]. The signals can be described reasonably well by three decaying oscillations with $v_1 \approx 0.52$ THz, $v_2 \approx 1$ THz, $v_3 \approx 1.53$ THz and $\gamma \approx 0.26$ ps$^{-1}$ (see Fig. S3). The fitted components are shown in the right panel of **Fig. 4a**. The dashed lines are extrapolations of the fitted curves beyond 8 ps. The change of the FID components over the course of 10 hours can be clearly seen. In this temporal window, the amplitude of the components $v_1 \approx 0.52$ THz and $v_3 \approx 1.53$ THz is reduced, whereas in contrast, the component $v_2 \approx 1$ THz gains larger amplitude. The amplitude variations, can also be seen in the Fourier spectra of the FID signals of the $H_2O@C_{60}$ in **Fig. 4b**, where the two ortho transitions lose their strength, while the para component becomes brighter. The temporal evolution of the latter amplitude variation, taken from the ortho component $v_1$ and the para component $v_2$, is given in **Fig. 4c**. Both curves can be fit well by single exponentials with ortho decay time of $11.22 \pm 0.14$ hrs and para rise time of $13.5 \pm 0.25$ hrs, in general agreement with previous findings [23,26]. We attribute the difference between the rise and decay time constants to the termination of the measurement prior to reaching the equilibrium state.

These results manifest the real-time conversion of the ortho-to-para spin isomers of water. Moreover, they demonstrate that the coherent rotation of entrapped water molecules lasts for about 10-15 ps inside its cage at 4 K. The damping of the rotational coherence may predominantly be attributed to the shaking of the $C_{60}$ cage due to lattice vibrations which couple to the rotational motion through translation-rotation coupling [24,29]. Since we used a diluted 5% $H_2O@C_{60}$ sample, water-water interactions may be ruled out as the damping mechanism of the rotational coherence.

It is important to note that, in the performed experiment, the measured decay rates correspond to the coherence relaxation time $T_2 = \gamma^{-1}$. Complementary information through the molecular alignment is required to obtain the population lifetime $T_1$ of the $H_2O@C_{60}$ rotational transitions [31]. Although, it can be acquired by time-resolved optical birefringence measurements for optically transparent samples, for the dark powders of $H_2O@C_{60}$ one may opt for the THz pump / THz probe experiment to determine $T_1$, an intriguing topic that we will address in a forthcoming publication.

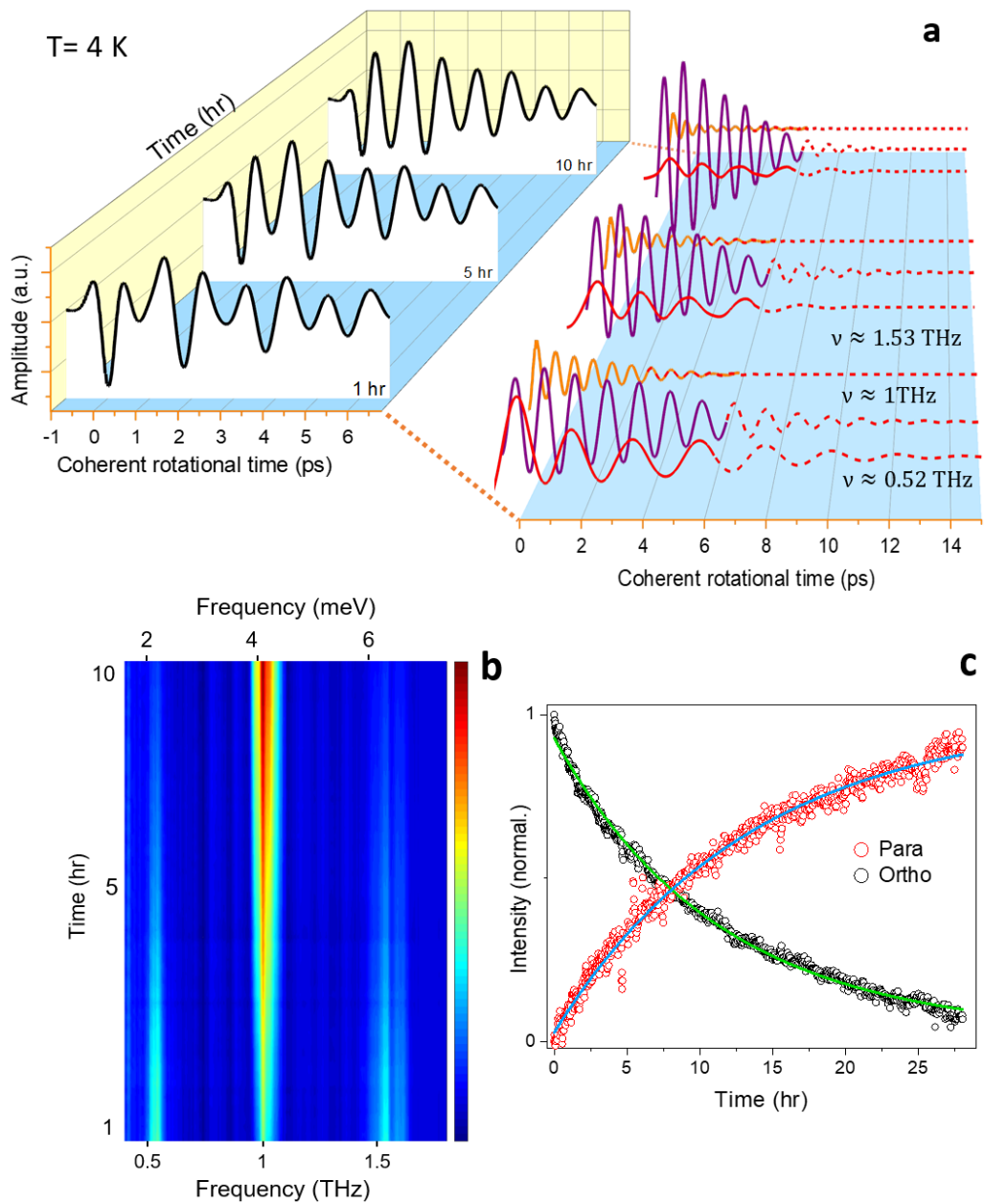

**Fig. 4**. **Ortho-para spin conversion of water at 4K**. **a**, The rotational motion of the encapsulated water in $C_{60}$ is measured as function of waiting time after a sudden cooling of $H_2O@C_{60}$ from 20 K to 4 K. The non-equilibrium population ratio of ortho and para spin isomers is directly observed by enhancing one oscillation at 1 THz and the reduction of frequency components at 1.53 THz and 0.52 THz. In the left panel, one observes the change of the pattern of the FIDs in the course of 10 hours. In the right panel, each trace is decomposed into its three frequency components. The dashed lines are extrapolations beyond 8 ps. **b**, The Fourier spectra of the time-domain rotational signal in panel a. In the time window of 10 hours the ortho lines at ~0.52 THz and ~1.53 THz lose their amplitude (coded dark blue), while the para line at ~1 THz gain larger amplitude (coded red). **c**, The kinetics of the amplitude of the para-water (@1THz) and ortho-water (@ 0.52 THz). The solid lines are single exponential fits to the experimental data (see text for details).

In conclusion, using single cycle THz pulses we launched a coherent wave-packet in the low-frequency rotational transitions of an ensemble of $H_2O@C_{60}$ (encapsulated water in fullerene-$C_{60}$) to resolve the emission of electromagnetic waves by the oriented dipoles. Underdamped coherent emissions are observed at temperatures below 100 K. At 4 K, the lowest temperature we reach, the rotation of water attains a long coherence decay time of about 10-15 ps. Although this is unusually long for molecular rotation in the solid state, it is still much shorter than that observed for water in the gaseous state at room temperature [36]. We also directly observe the change of the pattern of the dipole emission at 4 K, from a mixture of ortho- and para-water to a more purified para emission after a waiting period of 10s of hours, indicating the inter-conversion of spin isomers of water at cryogenic temperatures.

*Methods.*

*$H_2O@C_{60}$ sample.* 80% filled $H_2O@C_{60}$ was synthesized as described earlier [45], intimately mixed with 15 times its mass of commercial 99.5% $C_{60}$ (MTR Ltd) and the mixture sublimed in a quartz tube at $10^{-5}$ to $10^{-6}$ torr at 550 °C (with a 10°C/min temperature ramp). The product was assayed for composition by HPLC of a sample on a Cosmosil BuckyPrep analytical column (4.6 x 250mm) using toluene as the eluent at 1 ml/min which confirmed the desired 5% $H_2O$ filling.

*Time-domain THz spectroscopy.* We performed time-domain THz spectroscopy in the frequency interval of 0.15–2.5 THz using a commercial spectrometers TeraPulse and TeraView. The single cycle THz pulses were generate via photoconductive switching technique after excitation of the switches with femtosecond laser pulses from a Ti:Sapphire oscillator. Using a reverse process on a second photo-conductive switch we obtain the temporal profile of the THz pulses. The sample was cooled down to liquid helium temperatures in an Optistat (Oxford Instruments) helium-flow optical cryostat with Mylar windows. The THz pulse path was purged by nitrogen or vacuumed to exclude water vapor absorption lines. Each signal was obtained by averaging of about 200-500 pulses and accumulating for 30-40 seconds. The step size of the time-delay line movement provides a temporal resolution of 0.003 ps on the TeraView spectrometer and 0.008 ps on the TeraPulse spectrometer. In the temperature dependent measurements, the sample temperature was gradually lowered and the THz response of the system was measured at 10-degree steps and after 60 min waiting time. In addition to time-resolved experiments, frequency-dependent transmission coefficients of plane-parallel pressed pellets were recorded at fixed temperatures in the range 4 K-300 K.

**Acknowledgment**

We would like to thank A. Melentyev for his assistance in data processing and Toomas Rõõm, Stéphane Rols and Zlatko Bačić for fruitful discussions. The work in Stuttgart was supported by the Deutsche Forschungsgemeinschaft (DFG) via DR228/61-1. The work in Southampton was funded by the EPSRC-UK grants EP/P009980/1, EP/T004320/1 and EP/M001962/1.